\documentclass[prl,aps,showpacs,twocolumn,notitlepage]{revtex4-2}

\def\be{\begin{equation}}
\def\ee{\end{equation}}
\usepackage{graphicx}
\usepackage{bm}
\usepackage[latin9]{inputenc}
\usepackage{amsmath}
\usepackage{amssymb}
\usepackage{float}
\usepackage{lineno}
\usepackage[bookmarks=true,
   colorlinks=true,
   linkcolor=blue,
   urlcolor=blue,
   citecolor=blue,
   bookmarks=true,
   hyperindex=true
]{hyperref}
\usepackage{natbib}

\begin{document}

\title{Energy Dynamics of a Nonequilibrium Unitary Fermi Gas}

\author{Xiangchuan Yan$^{1}$}
\author{Jing Min$^{1,2}$}
\author{Dali Sun$^{1}$}
%
\author{Shi-Guo Peng$^{1}$}
\thanks{Corresponding author: pengshiguo@wipm.ac.cn}
\author{Xin Xie$^{1,2}$, Xizhi Wu$^{1,2}$}

\author{Kaijun Jiang$^{1,3}$}
\thanks{Corresponding author: kjjiang@wipm.ac.cn}

\affiliation{$^{1}$Innovation Academy for Precision Measurement Science and Technology, Chinese Academy of Sciences, Wuhan 430071, China}
\affiliation{$^{2}$University of Chinese Academy of Sciences, Beijing 100049, China}
\affiliation{$^{3}$Wuhan Institute of Quantum Technology, Wuhan 430206, China}

\date{\today}

\begin{abstract}
We investigate the energy dynamics of a unitary Fermi gas driven away from equilibrium. The energy is injected into the system by periodically modulating the trapping potential of a spherical unitary Fermi gas, and due to the existence of SO(2,1) symmetry, the breathing mode is excited without dissipation. Through the long-lived breathing oscillation, we precisely measure the energy evolution of the nonequilibrium system during the trap modulation. We find the trapping potential and internal energies increase with modulation time and simultaneously oscillate nearly $\textrm{180}^{\textrm{o}}$ out of phase. At large modulation amplitudes, the energy-injection efficiency is strongly reduced due to the trap anharmonicity. Unlike the equilibrium system, the measured energy evolution agrees well with predictions of the dynamic virial theorem. Our work provides valuable insights into the energy injection and redistribution in a non-equilibrium system, paving a way for future investigations of nonequilibrium thermodynamics.


\end{abstract}

\maketitle

Realization of a long-lived nonequilibrium system is crucial for exploring nonequilibrium dynamics \cite{Polkovnikov2011RMPnonequilibrium, eisert2015quantum, Gogolin_2016, Michael2018Nature}. Thanks to high controllability on interaction and geometry, ultracold atoms have offered a uniquely versatile platform for investigating the nonequilibrium properties \cite{langen2015ultracold, altman2015ARXIVnonequilibrium, Ueda2020NATUREREVIEWprethermalization, Chin2010RMPfeshbach} and probing  universal behaviors across various observables, such as spin correlation \cite {Oberthaler2018Nature}, momentum distribution \cite{Schmiedmayer2018Nature, Hadzibabic2018Nature} and Efimovian expansion \cite{Deng2016}. Energy is a fundamental quantity characterizing quantum matters. Exploring energy evolution in nonequilibrium
systems provides crucial insight into the thermodynamics of strongly interacting quantum gases \cite{Salomon2010NatureThermodynamics, Ueda2012ScienceThermodynamics, Zwierlein2012ScienceLamda}.

\begin{figure} [htbp]
\centerline{\includegraphics[width=8.5cm]{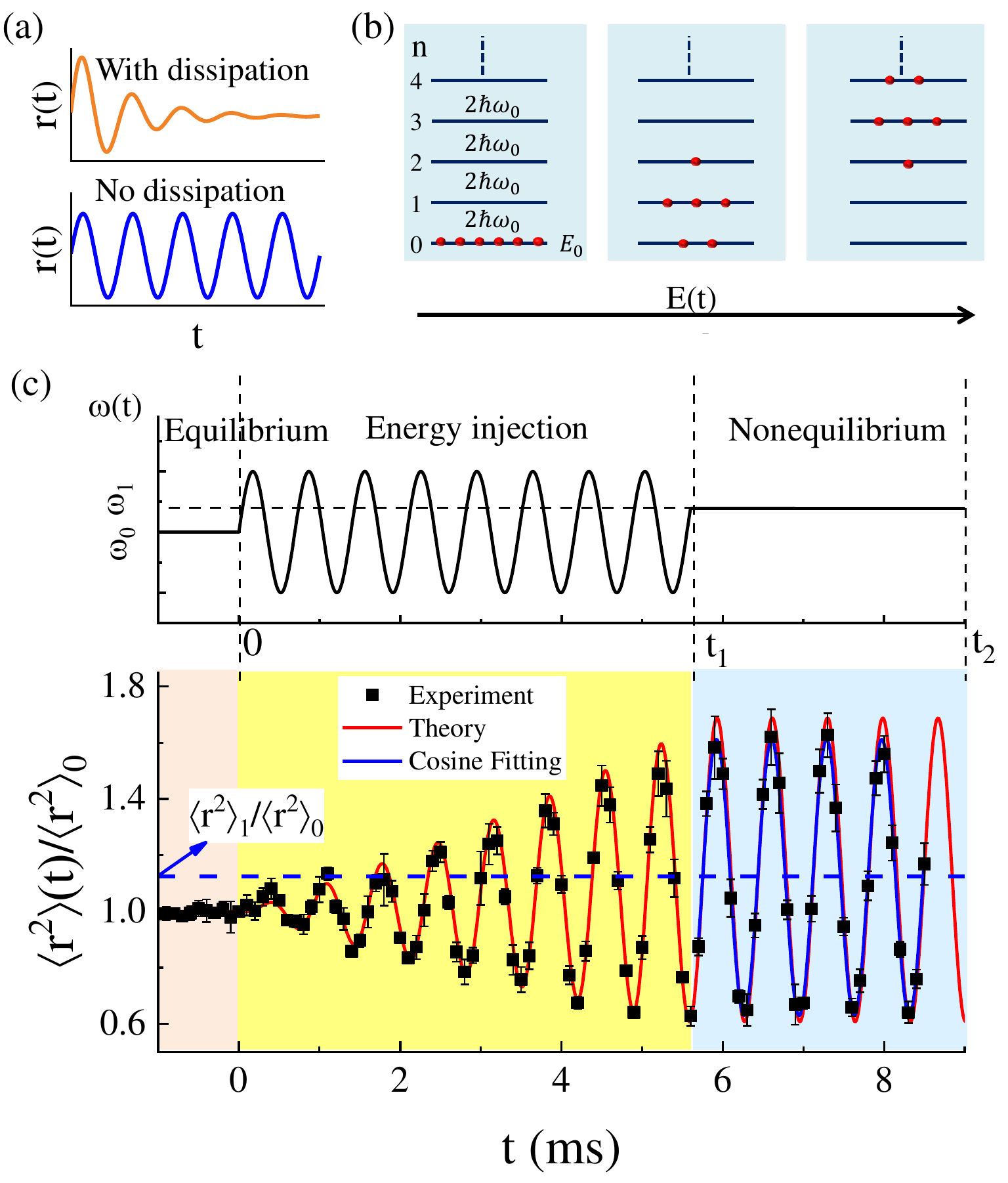}}
\caption{Measuring the energy injected into a nonequilibrium system. (a) Sketch of the atomic cloud size $r(t)$ after energy injection.  $r(t)$ will equilibrate to a constant value with the presence of dissipation (upper panel), while oscillates persistently without dissipation (lower panel). (b) The Fermi gas has conformal tower eigen states which have an equal separation of $2\hbar\omega_{0}$. As energy $E(t)$ is injected, higher-energy states are occupied, and the excited breathing mode remains in a coherent superposition state $\psi=\sum_{n}c_{n}\psi_{n}$. (c) Time sequence of the trap modulation (upper panel) and corresponsing atomic cloud sizes (lower panel). For $t\leq0$, the trapping frequency is set to $\omega_{0}$, keeping system in equilibrium.  For $0< t\leq t_{1}$, the trapping potential is isotropically modulated at frequency $2\omega_{0}$, injecting energy into the system. For $t> t_{1}$, the trapping frequency is fixed at $\omega_{1}$, where the excited breathing oscillation persists as a nonequilibrium state. The red curve represents the theoretical calculation of Eq. \eqref{eq:CloudSize}. Cloud sizes (black squares) for $t> t_{1}$ are fitted with a cosine function (blue solid curve), where $\left\langle r^{2}\right\rangle _{1}$ is the center position of the fitted oscillation (blue dashed line). The mean square cloud size $\left\langle r^{2}\right\rangle \left(t\right)$ is normalized to its equilibrium value $\left\langle r^{2}\right\rangle _{0}$. The modulation amplitude is $\beta=0.04$.} 
\label{fig:Fig1}
\end{figure}

For the unitary Fermi gas where zero-energy scattering length greatly exceeds interparticle spacing, the system features universal properties independent of interaction details. Energy could be injected into a system by rapidly modulating the trap \cite{science2005HeatThomas, PRL2005VirialThomas}. 
In most cases, dissipation occurs alongside energy injection, 
driving the system back to the equilibrium state after the trap modulation [see Fig. \ref{fig:Fig1}(a)]. The total energy of an equilibrium system is equal to twice the trapping potential energy, as predicted by the static virial theorem \cite{PRL2005VirialThomas, Thomas2008PRAvirial, werner2008PRAvirial, Castin2006PRAsymmety} and verified experimentally \cite{PRL2005VirialThomas, wang2024PRLscale}. In contrast, when there is no dissipation, 
the system will not equilibrate after the trap modulation, resulting in a long-lived nonequilibrium state \cite{Dalibard2002PRL, lobser2015observation, Vogt2012, sun2025PRApersistent}. Different from equilibrium systems, the energy of a nonequilibrium system is governed by the time-dependent collective motions, as predicted by the dynamic virial theorem  \cite{peng2023PRAvirial}. However, experimental observation of the energy dynamics in such nonequilibrium systems is lacking. 

A spherically trapped unitary Fermi gas has an undamped breathing mode due to the existence of SO(2,1) dynamical symmetry \cite{Pitaevskii1997PRAsymmetry, Castin2006PRAsymmety, wang2024PRLscale, sun2025PRApersistent}. During the excitation of breathing mode, the initial ground state $\psi_{0}$ is pumped to a coherent superposition state $\psi=\sum_{n}c_{n}\psi_{n}$ [see Fig. \ref{fig:Fig1}(b)], where $n$ is the quantum number of conformal tower states that have an equal separation of $2\hbar\omega_{0}$, and $c_{n}$ are expansion coefficients. With the increase of injected energy, higher energy states would be occupied, while dissipation is absent due to vanishing contribution of viscosities in a spherical unitary Fermi gas \cite{wang2024PRLscale, Maki2020, Maki2024PRA, Ho2004, Son2007}. Furthermore, the long-lived characteristic of the breathing oscillation enables us to precisely measure the energy of the nonequilibrium system, as predicted by the dynamic virial theorem \cite{peng2023PRAvirial, SM2024}.

In this work, we systematically study the energy dynamics of a nonequilibrium unitary Fermi gas. By periodically modulating the trapping potential of a spherical unitary Fermi gas, we inject energy into system without inducing dissipation. The long-lived breathing mode excited allows us to precisely measure the energy evolution during the modulation process, providing direct insight into the energy dynamics of a nonequilibrium system. We further measure the trapping potential energy and internal energy, respectively, and observe that the two energy components increase with modulation time and simultaneously oscillate nearly $\textrm{180}^{\textrm{o}}$ out of phase. At large modulation amplitudes beyond harmonic limit, the energy-injection efficiency is strongly reduced due to the anharmonicity of the trap. Notably, the experimental results for the injected energy are in good agreements with predictions of the dynamic virial theorem, different from equilibrium systems. 

The time-dependent energy $E\left(t\right)$ of a unitary Fermi gas is constrained by the dynamic virial theorem \cite{peng2023PRAvirial, SM2024}
\begin{equation}
E\left(t\right)=2E_{\text{ho}}\left(t\right)+\frac{1}{4}\frac{d^{2}I(t)}{dt^{2}},\label{eq:DynamicVirialTheorem}
\end{equation}
which reveals the energy dynamics of strongly-interacting many-body systems. Here, $E_{\text{ho}}=\left\langle m\omega^{2}r^{2}(t)\right\rangle/2 $ is the trapping potential energy, and $I(t)=\left\langle mr^{2}(t)\right\rangle $ is the moment of inertia, with $m$ being the atomic mass and $\omega$ the trapping frequency. Equation \eqref{eq:DynamicVirialTheorem} constructs a dependence of the instantaneous energy $E\left(t\right)$ on mean square cloud size $\left\langle r^{2}\right\rangle \left(t\right)$. In the equilibrium limit, it reduces to the static virial theorem, i.e., $E=2H_{ho}=m\omega^{2}\left\langle r^{2}\right\rangle$ \cite{PRL2005VirialThomas, Thomas2008PRAvirial, tan2008ANNALSvirial}.  

In the experiment, a degenerate $^{6}$Li atomic Fermi gas with two balanced spin states $\left|F=1/2,\ m_{F}=\pm1/2\right\rangle $ is prepared in a spherical trap as in the previous works \cite{wang2024PRLscale, sun2025PRApersistent}. The magnetic field is at $B=832.2$ G, right on the Freshbach resonance. The total atomic number is $N=9.6\times10^{4}$, the temperature is $T/T_{F}=0.24$ ($T_{F}$ is the Fermi temperature of a noninteracting gas), and the trapping frequency is $\omega_{0}=2\pi\times720$ Hz with a small asphericity of about $5\%$ \cite{Asphericity}. 
The experimental progress is schematically shown in Fig. \ref{fig:Fig1}(c). The system is initially prepared in equilibrium ($t\leq0$). Then the trapping potential is isotropically modulated at a frequency of $2\omega_{0}$ for a duration $t_{1}$, i.e., $0< t\leq t_{1}$, injecting energy into the system.
The modulation follows $\omega^{2}\left(t\right)=\omega_{0}^{2}\left[1+\beta\sin\left(2\omega_{0}t\right)\right]$ with modulation amplitude $\beta$. After modulation ($t>t_{1}$), the trap is held constant at $\omega_{1}$, and the atomic cloud size exhibits long-lived oscillates, forming a nonequilibrium quantum state. 

\begin{figure} [htbp]
\centerline{\includegraphics[width=8cm]{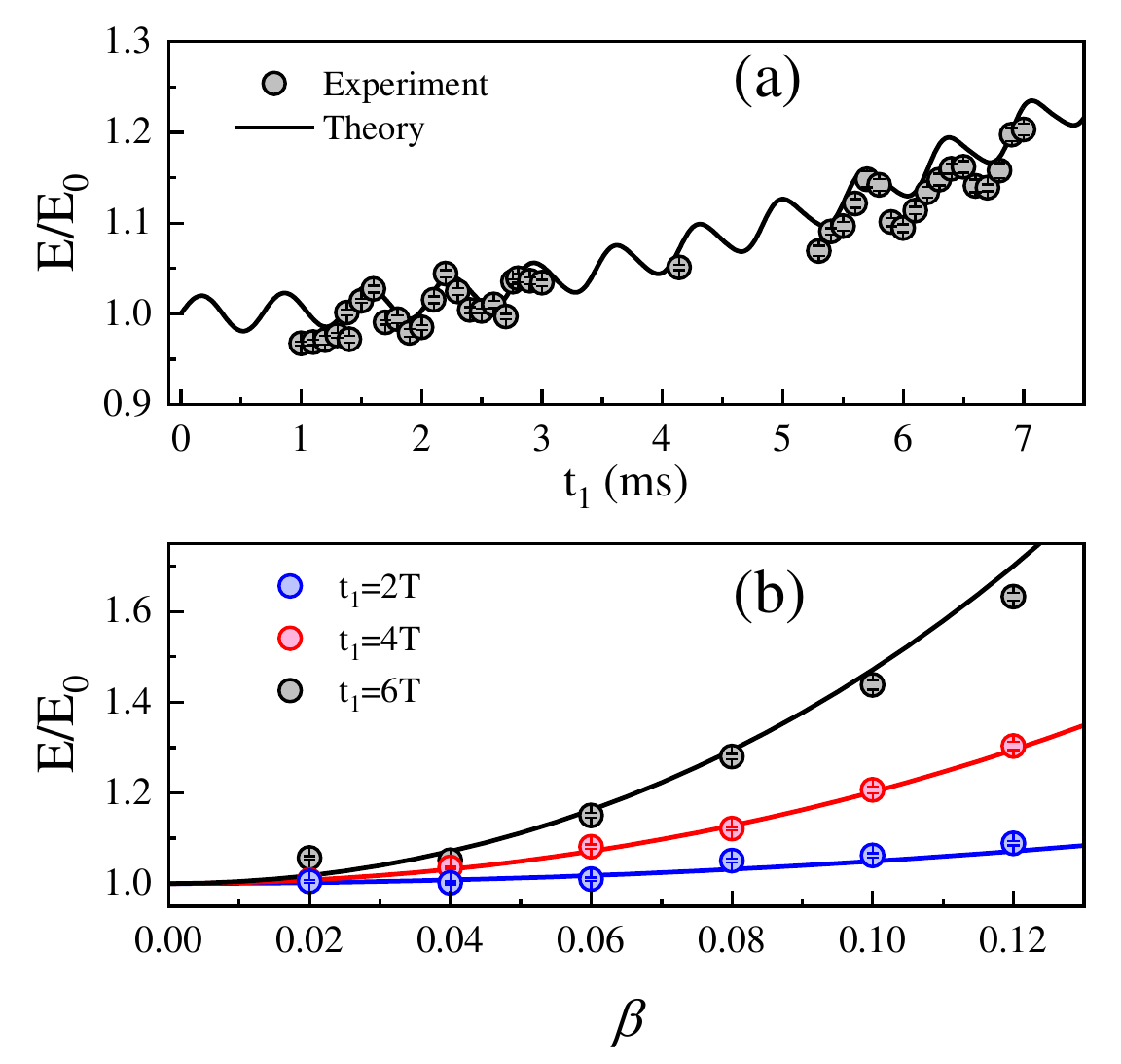}}
\caption{Energy evolution during the trap modulation. (a) Energy $E$ as a function of the modulation time $t_{1}$. The black solid curve denotes the calculation of Eq. \eqref{eq:EnergyEvolution}. The modulation amplitude is $\beta=0.04$. $E$ is normalized to the initial equilibrium energy $E_{0}$. The error bars, representing the standard deviations of several measurements, are smaller than the marker size. (b) Energy $E$ versus the excitation amplitude $\beta$. The modulation time $t_{1}$ is set as $2T$, $4T$ and $6T$, respectively, where $T=\pi/\omega_{0}$ is the modulation period. The three solid curves denote the calculations of Eq. \eqref{eq:EnergyEvolution}.}
\label{fig:Fig2}
\end{figure}

For $t>t_{1}$, it exhibits an undamped breathing oscillation at twice the trapping frequency due to the existence of SO$\left(2,1\right)$ symmetry \cite{wang2024PRLscale, sun2025PRApersistent, peng2023PRAvirial, Pitaevskii1997PRAsymmetry, Castin2006PRAsymmety}, i.e., $\left\langle r^{2}\right\rangle \left(t\right)=E/m\omega_{1}^{2}+A\cos(2\omega_{1}t+\delta)$, where $\omega_{1}$ is the trapping frequency, and $A$ and $\delta$ are the oscillation amplitude and phase, respectively. We fit the measured atomic cloud sizes using a cosine function [blue solid curve in Fig. \ref{fig:Fig1}(c)], and the center position of the fitted oscillation [blue dashed curve in Fig. \ref{fig:Fig1}(c)] relates the energy of the nonequilibrium system, i.e., $\left\langle r^{2}\right\rangle _{1}=E/m\omega_{1}^{2}$. 
It is noted that the atomic cloud is probed after a $1$ ms time-of-flight (TOF) expansion. In this condition, the center position relates the energy $\left(1+\omega_{1}^{2}t_{\text{TOF}}^{2}\right)E/m\omega_{1}^{2}$ \cite{SM2024}. Using this method, we can measure the energy evolution during the trap modulation, which are shown in Fig. \ref{fig:Fig2}.  

For a spherically trapped unitary Fermi gas, the system exhibits scale invariance \cite{wang2024PRLscale}, which imposes an exact constraint on the evolution of the cloud size \cite{SM2024, Stringari2002PRLexpansion, Hu2004PRLexpansion}
\begin{equation}
\frac{d^{2}b(t)}{dt^{2}}+\omega^{2}\left(t\right)b\left(t\right)-\frac{\omega_{0}^{2}}{b^{3}\left(t\right)}=0,\label{eq:CloudSize}
\end{equation}
where $b\left(t\right)$ represents the ratio of the instantaneous cloud size $\left\langle r\right\rangle \left(t\right)$ to its initial equilibrium size $\left\langle r\right\rangle _{0}$. The evolution of cloud sizes calculated with Eq. \eqref{eq:CloudSize} is displayed in Fig. \ref{fig:Fig1}(c). 
The oscillation amplitude of the cloud size increases with modulation time, indicating continuous injection of energy into the system. Once the modulation stops, the system exhibits long-lived breathing oscillations, sustained by the SO(2,1) symmetry. 

From the dynamic virial theorem Eq. \eqref{eq:DynamicVirialTheorem}, the energy of the system during the modulation is calculated from the instantaneous cloud size as \cite{SM2024}
\begin{equation}
\frac{E\left(t\right)}{E_{0}}=\frac{1}{4\omega_{0}^{2}}\frac{d^{2}b^{2}(t)}{dt^{2}}+\frac{\omega^{2}\left(t\right)}{\omega_{0}^{2}}b^{2}\left(t\right),\label{eq:EnergyEvolution}
\end{equation}
where $b\left(t\right)$ is determined by Eq.(\ref{eq:CloudSize}), and $E_{0}$ is the energy of the initial equilibrium state.
The calculation of energy evolution during the trap modulation, as shown in Fig. \ref{fig:Fig2}, agrees well with experimental measurements. Despite relative small injected energy, the energy evolution can be clearly resolved.  
As illustrated in Fig. \ref{fig:Fig2}(a), the energy $E$ increases with modulation time $t_{1}$ and oscillates synchronously with the trap modulation. This behavior is expected, since energy is injected when the trap is compressed and partially released when it expands. In addition, the injected energy monotonically increases with the modulation amplitude $\beta$, as shown in Fig. \ref{fig:Fig2}(b). Overall, we can precisely control the injected energy by adjusting the modulation time $t_{1}$ or amplitude $\beta$.

\begin{figure}[htbp]
\centerline{\includegraphics[width=8cm]{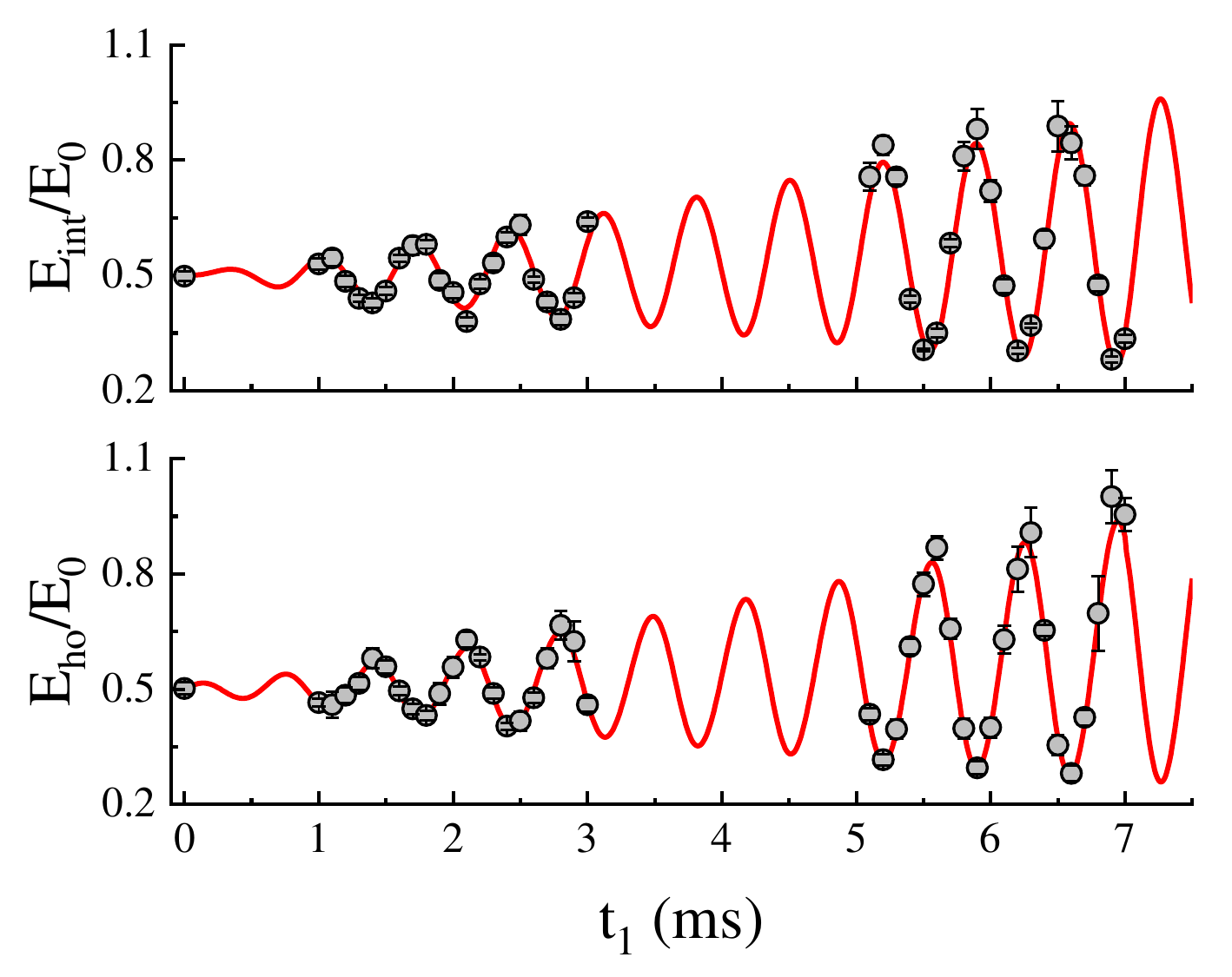}}
\caption{Distribution of two energy components during the trap modulation. The internal energy $E_{\text{int}}/E_{0}$ and trapping potential energy $E_{\text{ho}}/E_{0}$ are shown on the upper and lower panels, respectively. The red solid curves represent calculations of Eq. \eqref{eq:EnergyEvolution}. Each of two energy components increases and oscillates with the modulation time $t_{1}$. Here, $E_{0}$ is the initial equilibrium energy, and the modulation amplitude is $\beta=0.04$. Error bar is the standard deviation of several measurements.
}
\label{fig:Fig3}
\end{figure}

The total energy $E\left(t\right)$ consists of external trapping potential energy $E_{\text{ho}}\left(t\right)$ and internal energy $E_{\text{int}}\left(t\right)$.  To investigate the energy distribution during the trap manipulation, we could measure different energy components. The internal energy is equal to the sum of kinetic and interaction energies. When the trap is switched off immediately after the trap modulation, interaction energy gradually converts into kinetic energy during the atomic expansion. In the long-time expansion limit, internal energy $E_{\text{int}}$ is equal to kinetic energy and could be extracted by measuring the expansion velocity. Meanwhile, the trapping potential energy $E_{\text{ho}}\left(t\right)$ is related to the in-situ cloud size and could be deduced from the size measured after 1 ms TOF \cite{SM2024}. The experimental results of internal energy $E_{\text{int}}$ and trapping potential energy $E_{\text{ho}}\left(t\right)$, as shown in Fig. \ref{fig:Fig3}, are in good agreements with calculations of Eq. \eqref{eq:EnergyEvolution}.  
The two energy components increase with modulation time and simultaneously oscillate nearly $\textrm{180}^{\textrm{o}}$ out of phase, conversing between each other. To understand this behavior of the energy redistribution, we consider a modulation with a small amplitude ($\beta\ll1$) in a short time. Linearizing Eq. (\ref{eq:CloudSize}) by setting $b\left(t\right)\approx1+\epsilon\left(t\right)$ with $\epsilon\left(t\right)\ll1$, we  solve the linearized equation and obtain $\epsilon\left(t\right)$, which yields \cite{SM2024} 
\begin{equation} \label{eq:TwoComponent}
\begin{split}
\frac{E_{\text{ho}}\left(t\right)}{E_{0}} & \approx  \frac{1}{2}+\frac{\beta}{4}\left[\frac{3}{2}\sin\left(2\omega_{0}t\right)+\omega_{0}t\cos\left(2\omega_{0}t\right)\right],\\
\frac{E_{\text{int}}\left(t\right)}{E_{0}} & \approx  \frac{1}{2}+\frac{\beta}{4}\left[\frac{1}{2}\sin\left(2\omega_{0}t\right)-\omega_{0}t\cos\left(2\omega_{0}t\right)\right]
\end{split}
\end{equation}
up to the first-order term in $\beta$. For $\omega_{0}t\gg3/2$ (or t$\gg0.3$ ms), a condition well satisfied within the timescale of the experiment, the cosine terms dominate in the variation of energy, and Eq. \eqref{eq:TwoComponent} clearly indicates that $E_{\text{ho}}\left(t\right)$ and $E_{\text{int}}\left(t\right)$ oscillate nearly $\textrm{180}^{\textrm{o}}$ out of phase.

Certainly, summing $E_{\text{ho}}\left(t\right)$ and $E_{\text{int}}\left(t\right)$ can also enable us to obtain the total energy $E\left(t\right)$ during the trap modulation. Using this method, the measured energy is consistent with the calculation of Eq. \eqref{eq:EnergyEvolution}, while it has much larger fluctuation than the measurement of breathing oscillation (see Supplemental Materials \cite{SM2024}). This indicates that the long-lived characteristic of breathing mode greatly improves the accuracy of energy measurement.

\begin{figure}
\centerline{\includegraphics[width=8cm]{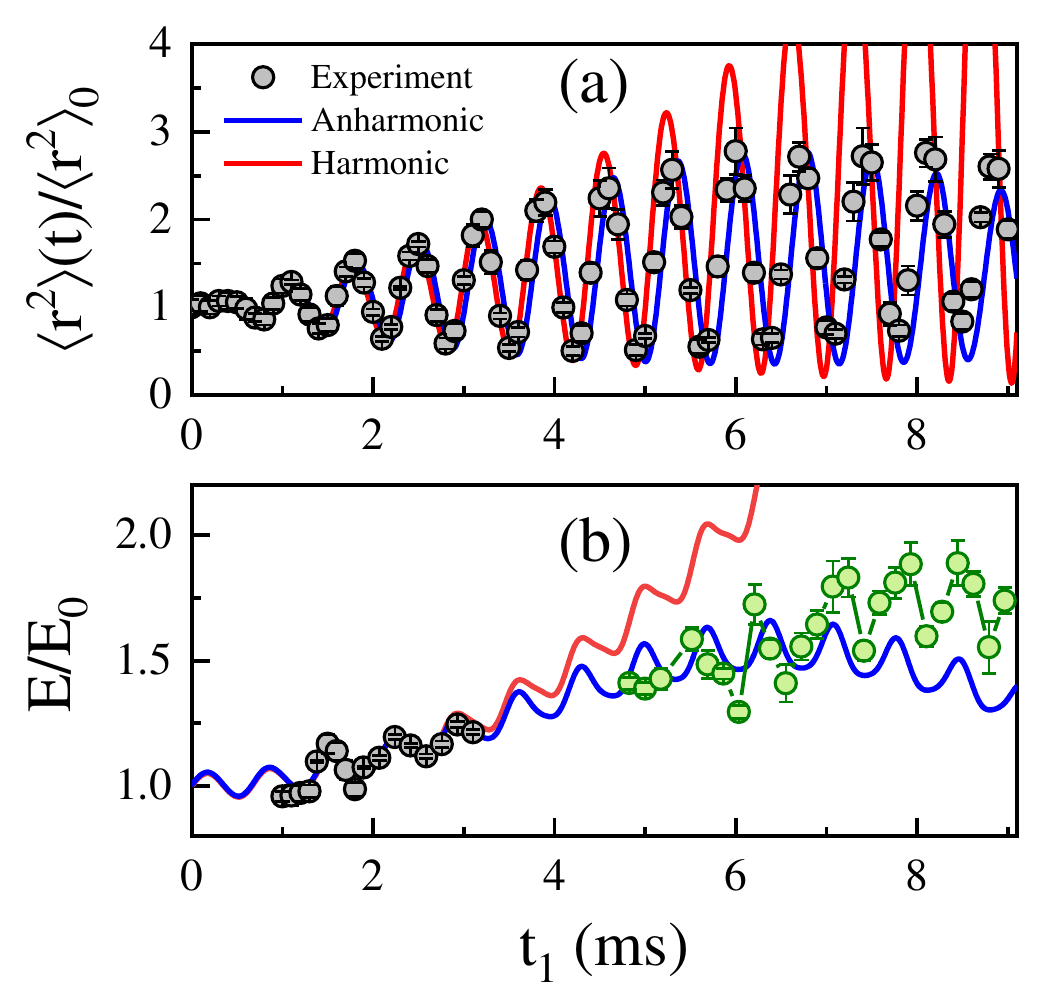}}
\caption{Injected energy at a large modulation amplitude. The modulation amplitude is $\beta=0.1$. (a) Evolution of cloud sizes versus the modulation time $t_{1}$. The red solid curve represents the calculation of Eq. \eqref{eq:CloudSize} in a harmonic trap, while the blue solid curve includes the anharmonicity correction using Eq. \eqref{eq:Vext}. (b) Evolution of energy versus the modulation time $t_{1}$. The energy is calculated with Eq. \eqref{eq:EnergyEvolution} in a harmonic trap (red solid curve) and with Eq. \eqref{eq:DVT_anho} when considering anharmonicity (blue solid curve). The injected energies for short modulation times are measured from the long-lived breathing oscillation (black circles), whereas those for long modulation times are obtained from TOF expansion (green circles). Error bars indicate the standard deviation of several measurements. Here, $\left\langle r^{2}\right\rangle _{0}$ and $E_{0}$ denote mean square cloud size and energy of the initial equilibrium system, respectively. 
}
\label{fig:Fig4}
\end{figure}

When the modulation amplitude increases or the modulation time extends, both the oscillation amplitude of the cloud size and the injected energy grow more slowly than predicted in the harmonic trap, as illustrated
in Fig. \ref{fig:Fig4}. This deviation arises from the trap anharmonicity, which becomes increasingly large as the atomic cloud size grows during modulation. 
Expanding the trapping potential $V_{\text{ext}}\left({\bf r}\right)$ around the trap center (${\bf r}=0$) up to the fourth order gives \cite{SM2024}
\begin{equation}
V_{\text{ext}}\left({\bf r},t\right)\approx\frac{1}{2}\omega^{2}\left(t\right)\left\{ r^{2}-\frac{\eta}{2}\left[r^{4}+\left(x^{2}-y^{2}\right)^{2}\right]\right\} ,\label{eq:Vext}
\end{equation}
where $\omega_{0}=\sqrt{4U_{0}/mw^{2}}$ is the harmonic trapping frequency with the laser beam waist $w$, and $a_{\text{ho}}=\sqrt{\hbar/m\omega_{0}}$ is the harmonic length. The dimensionless parameter $\eta=\left(a_{\text{ho}}/w\right)^{2}$ quantifies the anharmonicity of the trap. 
The anharmonicity modifies the eigenfrequencies of the collective modes. Within the framework of hydrodynamic theory, the frequency of the breathing mode is shifted to $\omega_{B}/\omega_{0}=2-19\eta\chi_{0}/10$, where $\chi_{0}$ is the initial value of $\chi\left(t\right)\equiv\left\langle r^{4}\right\rangle \left(t\right)/a_{\text{ho}}^{2}\left\langle r^{2}\right\rangle \left(t\right)$ at equilibrium. In our experiment at an initial temperature $T/T_{F}=0.24$, the anharmonicity parameter is $\eta\approx4.6\times10^{-4}$ with $\chi_{0}\approx130$. As shown in Fig. \ref{fig:Fig4}(a), for a modulation amplitude as large as  $\beta=0.1$, the oscillation amplitude of cloud sizes measured at long modulation times is significantly smaller than the calculation of Eq. \eqref{eq:CloudSize} in a harmonic trap. However, it agrees well with the theoretical prediction when considering anharmonicity in Eq. \eqref{eq:Vext}. The anharmonicity causes a detuning of the modulation from the resonant frequency, 
leading to a slower increase in the oscillation amplitude of cloud size.

We develop the dynamic virial theorem including the trap anharmonicity \cite{SM2024}
\begin{equation}
E\left(t\right)=\frac{1}{4}\frac{d^{2}I(t)}{dt^{2}}+\left\langle V_{\text{ext}}\right\rangle +\frac{1}{2}\left\langle {\bf r}\cdot \nabla V_{\text{ext}}\right\rangle ,\label{eq:DVT_anho}
\end{equation}
which reduces to the standard form at equilibrium \cite{Thomas2008PRAvirial}, $E=\left\langle V_{\text{ext}}\right\rangle +\left\langle {\bf r}\cdot\nabla V_{\text{ext}}\right\rangle /2$. For short modulation times ($t_{1}<4 \ \textrm{ms}$), the anharmonicity is small, allowing energy to be extracted by probing the long-lived breathing oscillation [Fig. \ref{fig:Fig4}(b)].  
As the modulation extending, the trap anharmonicity increases to break the SO(2,1) symmetry, which introduces damping in breathing oscillation and eventually drives the system toward a new equilibrium. 
To abtain the injected energy in this case, atoms are allowed to freely oscillate until equilibrium is reached, followed by a TOF expansion. 
The internal energy, $E_{\text{int}}=\left\langle {\bf r}\cdot\nabla V_{\text{ext}}\right\rangle /2$, is obtained by measuring the expanding velocity of atomic cloud in the long-time expansion. From the explicit form of $V_{\text{ext}}\left({\bf r}\right)$, we obtain $E\approx\left(2+19\eta\chi/30\right)E_{\text{int}}$ \citep{SM2024}, where $\chi$ depends on the final equilibrium temperature. For long modulation times ($t_{1}>4 \ \textrm{ms}$), the measured energies are much smaller than the calculations of Eq. \eqref{eq:EnergyEvolution} in a harmonic trap [see Fig. \ref{fig:Fig4}(b)]. When considering anharmonicity via Eq. \eqref{eq:DVT_anho}, the experimental results show reasonable agreements with the theoretical calculations. The slightly larger values in  measurements probably arises from heating effects during the equilibrating process.

In conclusion, we report a novel method to explore the energy evolution of a nonequilibrium quantum gas. The SO(2,1) dynamical symmetry of the spherically trapped unitary Fermi gas provides a valid platform for energy injection without dissipation and production of an long-lived breathing oscillation. The energy evolution in the nonequilibrium Fermi gas is in good agreement with predictions of the dynamic virial theorem, different from the equilibrium system. The role of the trap anharmonity on the energy injection at large modulation amplitudes has been clarified in both experiment and theory. Our work achieves time-resolved observation of energy evolution, providing valuable insights into energy injection and redistribution in a nonequilibrium quantum system.



Our study primarily focuses on the unitary regime with a divergent $s$-wave scattering length ($a\rightarrow\infty$). Away from the resonant interaction, the dynamic virial theorem in Eq. \eqref{eq:DynamicVirialTheorem} extends to a more general form that includes time-dependent scattering length and Tan's contact \cite{peng2023PRAvirial}.
Our system, spanning the BEC-BCS crossover, provides a promising avenue for probing the energy of non-equilibrium quantum gases at finite interactions. Future studies may further extend this approach to measure the energy of many-body quantum gases under interaction modulation \cite{Zhai2021PRLmaximumEnergy, Shi2021PRAtimedependent, Maki2022PRLtimedependent} or quench dynamics \cite{Maki2020, Aditi2018AnnualQuench}.  

We thank Shizhong Zhang for favorit discussions. This work is supported by the National Key R\&D Program under Grant No. 2022YFA1404102, NSFC (Grants No. 12374250, No. U23A2073 and No. 12121004), Innovation Program for Quantum Science and Technology under Grant No. 2023ZD0300401, the Natural Science Foundation of Hubei province under Grant No. 2021CFA027 and Wuhan city under Grant No. 2024040701010063. 

X.Y, J.M and D.S contributed equally to this work.



%

\newpage
\begin{widetext}

\setcounter{secnumdepth}{3} 

\setcounter{equation}{0}

\setcounter{figure}{0}
	
\renewcommand{\thefigure}{S\arabic{figure}}
\renewcommand{\thetable}{S\arabic{table}}
\renewcommand{\theequation}{S\arabic{equation}}

\section*{Supplemental materials}

\section{Atomic cloud size during the trap manipulation}
Let us consider the dynamics of a unitary Fermi gas within the framework of hydrodynamic theory \cite{Thomas2014PRLconformalSpp}. The governing equations include the continuity equation:
\begin{equation}
\frac{\partial n}{\partial t}+\nabla\cdot\left(n{\bf v}\right)=0,\label{eq:theory1}
\end{equation}
and the Euler equation 
\begin{equation}
\left[n\frac{\partial}{\partial t}+n\left({\bf v}\cdot\nabla\right)\right]v_{i}=-\frac{1}{m}\frac{\partial p}{\partial r_{i}}-\frac{n}{m}\frac{\partial V_{\text{ext}}}{\partial r_{i}}+\frac{1}{m}\sum_{j}\frac{\partial\left(\eta\sigma_{ij}\right)}{\partial r_{j}},\label{eq:theory2}
\end{equation}
where $n\left({\bf r},t\right)$ is the atom density, ${\bf v}\left({\bf r},t\right)$ is the velocity field, $p\left({\bf r},t\right)$ is the local pressure, $V_{\text{ext}}\left({\bf r},t\right)$ is the external potential, and $\sigma_{ij}=\partial v_{i}/\partial r_{j}+\partial v_{j}/\partial r_{i}-2\delta_{ij}\nabla\cdot{\bf v}/3$ is the shear tress tensor. The shear viscosity is denoted by $\eta$, and for a unitary Fermi gas, the bulk viscosity vanishes.

To analyze the evolution of the atomic cloud size $\left\langle r_{i}^{2}\right\rangle $, we adopt a scaling ansatz for the density profile at time $t$ \cite{Stringari2002PRLexpansionSpp, Hu2004PRLexpansionSpp}
\begin{equation}
n\left(x,y,z,t\right)=\frac{1}{b_{x}b_{y}b_{z}}n_{0}\left(\frac{x}{b_{x}},\frac{y}{b_{y}},\frac{z}{b_{z}}\right),\label{eq:theory3}
\end{equation}
which relates the time-dependent density profile $n\left({\bf r},t\right)$ to the equilibrium density profile $n_{0}\left({\bf r}\right)$ by scaling factors $b_{i}\left(t\right)$ for $i=x,y,z$. Then the time dependence of the could size is entirely captured by scaling factors $b_{x,y,z}$, which satisfy
\begin{equation}
\frac{d^{2}b_{i}}{dt^{2}}-\frac{b_{i}^{-1}\left(b_{x}b_{y}b_{z}\right)^{-2/3}}{m\left\langle r_{i}^{2}\right\rangle _{0}}\left\langle r_{i}\frac{\partial V_{\text{ext}}}{\partial r_{i}}\right\rangle _{0}+\frac{b_{i}^{-1}}{m\left\langle r_{i}^{2}\right\rangle _{0}}\left\langle r_{i}\frac{\partial V_{\text{ext}}}{\partial r_{i}}\right\rangle _{t}=0,\label{eq:theory4}
\end{equation}
and $\left\langle \right\rangle _{0,t}$ denotes the values at equilibrium and at time $t$. For an isotropic harmonic trap, where $b_{x}=b_{y}=b_{z}\equiv b$, we find that the scaling factor $b$ satisfies the equation \cite{CASTIN2004407Spp,PhysRevA.74.053604Spp}
\begin{equation}
\ddot{b}\left(t\right)+\omega^{2}\left(t\right)b\left(t\right)-\frac{\omega_{0}^{2}}{b^{3}\left(t\right)}=0.\label{eq:theory5}
\end{equation}
Regarding the modulation of the trapping potential $U(t)$ in our experiment (i.e., $U(t)\propto \omega^{2}(t)$), the time-dependent trapping frequency follows the form of 
\begin{equation}
\omega^{2}(t)=\left\{ \begin{array}{ll}
\omega_{0}^{2}, & t\leq0,\\
\omega_{0}^{2}\left[1+\beta\sin\left(2\omega_{0}t\right)\right], & 0<t\leq t_{1},\\
\omega_{1}^{2}=\omega_{0}^{2}\left[1+\beta\sin\left(2\omega_{0}t_{1}\right)\right], & t_{1}<t\le t_{2}.
\end{array}\right.\label{eq:theory6}
\end{equation}
In this scenario, we initially prepare a unitary Fermi gas at equilibrium in the trap with frequency $\omega_{0}$. At time $t=0$, we begin sinusoidal modulation of the trapping potential with the modulation
frequency $2\omega_{0}$ and amplitude $\beta$. At time $t_{1}$, we stop the modulation and maintain a constant trapping frequency of $\omega_{1}$, allowing the atomic cloud to undergo breathing-mode oscillations. The time evolution of the cloud size is calculated with Eq. \eqref{eq:theory5}.

\section{energy evolution during the trap manipulation}

In this section, we derive the time-dependent total energy during the modulation process $0<t\le t_{1}$. To calculate the energy dynamics of a unitary Fermi gas, we apply the dynamic virial theorem as \cite{peng2023PRAvirialSpp}
\begin{equation}
E\left(t\right)=\frac{1}{4}\frac{d^{2}I(t)}{dt^{2}}+2E_{\text{ho}}\left(t\right),\label{eq:theory7}
\end{equation}
where $E_{\text{ho}}\left(t\right)$ is the potential energy and $I\left(t\right)$ is the moment of inertia, both dependent on the mean square size $\left\langle r^{2}\right\rangle \left(t\right)$. Diving both sides of Eq. \eqref{eq:theory7} by the initial energy at equilibrium, $E_{0}=m\omega_{0}^{2}\left\langle r^{2}\right\rangle _{0}$, we obtain 
\begin{equation}
\frac{E\left(t\right)}{E_{0}}=\frac{1}{4\omega_{0}^{2}}\frac{d^{2}b^{2}(t)}{dt^{2}}+\frac{\omega^{2}\left(t\right)}{\omega_{0}^{2}}b^{2}\left(t\right),\label{eq:theory8}
\end{equation}
where the scaling factor $b\left(t\right)$ satisfies Eq. \eqref{eq:theory5} with $\omega\left(t\right)=\omega_{0}^{2}\left[1+\beta\sin\left(2\omega_{0}t\right)\right]$ during the modulation. The energy evolution during the modulation is then calculated with Eq. \eqref{eq:theory8}. 

\section{Relation between energy and center position of the oscillation}

At time $t_{1}$, we stop the trap modulation and maintain the trapping frequency constant at $\omega_{1}=\omega_{0}^{2}\left[1+\beta\sin\left(2\omega_{0}t_{1}\right)\right]$, allowing the atomic cloud to freely oscillate. The total energy is conserved during this free oscillation and equals to the energy at the end of the modulation, $E=E\left(t_{1}\right)$ for $t_{1}<t\le t_{2}$. 
The dynamic virial theorem predicts an oscillation of the cloud size, 
\begin{equation}
\begin{aligned}\left\langle r^{2}\right\rangle \left(t\right)\end{aligned}
=A\cos\left(2\omega_{1}t+\delta\right)+\frac{E}{m\omega_{1}^{2}},\label{eq:theory9}
\end{equation}
where $A$ is the oscillation amplitude and $\delta$ is the phase, both determined by the initial conditions at $t=t_{1}$. Then the total energy $E$ is directly related to the center position of the oscillation $\left\langle r^{2}\right\rangle _{1}$,  i.e., $\left\langle r^{2}\right\rangle _{1}=E/m\omega_{1}^{2}$. However, due to the small in-situ size of the atomic cloud, we turn off the trapping potential at $t=t_{2}$ and probe the cloud size after a time of flight (TOF) $t_{\text{TOF}}$. 

During the free oscillation in the trap, i.e., $t_{1}<t\le t_{2}$, the evolution of the cloud size is governed by Eq. \eqref{eq:theory9}. At the time of switching off the trap, $t=t_{2}$, we have

\begin{align}
\left\langle r^{2}\right\rangle \left(t_{2}\right) & =A\cos\left(2\omega_{1}t_{2}+\delta\right)+\frac{E}{m\omega_{1}^{2}},\label{eq:theory10}\\
\frac{d\left\langle r^{2}\right\rangle \left(t_{2}\right)}{dt} & =-2\omega_{1}A\sin\left(2\omega_{1}t_{2}+\delta\right),\label{eq:theory11}
\end{align}
and the internal energy is
\begin{equation}
E\left(t_{2}\right)=E-\frac{1}{2}m\omega_{1}^{2}\left\langle r^{2}\right\rangle \left(t_{2}\right)=\frac{1}{2}\left[E-Am\omega_{1}^{2}\cos\left(2\omega_{1}t_{2}+\delta\right)\right].\label{eq:theory12}
\end{equation}
By solving Eq. \eqref{eq:theory5} with initial conditions of Eqs. \eqref{eq:theory10} and \eqref{eq:theory11}, we obtain the cloud size after a TOF of $t_{\text{TOF}}$ as a function of the holding time $t_{2}$,
\begin{equation}
\left\langle r^{2}\right\rangle \left(t_{2}\right)=\left(1-\omega_{1}^{2}t_{\text{TOF}}^{2}\right)A\cos\left(2\omega_{1}t_{2}+\delta\right)-2\omega_{1}t_{\text{TOF}}A\sin\left(2\omega_{1}t_{2}+\delta\right)+
\left(1+\omega_{1}^{2}t_{\text{TOF}}^{2}\right)\frac{E}{m\omega_{1}^{2}}.\label{eq:theory13}
\end{equation}
Equation \eqref{eq:theory13} indicates that the center position of the oscillation after a TOF equals to the energy $\left(1+\omega_{1}^{2}t_{\text{TOF}}^{2}\right)E/m\omega_{1}^{2}$. 

\section{Measurement of different energy components during the manipulation}

The energy injected into the system distribute among different components, such as the external trapping potential energy $E_{\text{ho}}$ and internal $E_{\text{int}}$. We could measure the distribution of different energy components during the trap modulation. The internal energy equals to the sum of kinetic and interaction energies. In Fig. \ref{fig:sFig1}, we give three examples to demonstrate how to measure $E_{\text{int}}$ and $E_{\text{ho}}$, respectively. When switching off the trap at $t=t_{1}$ immediately after the trap modulation, interaction energy is gradually converted into kinetic energy during the atomic expansion. The internal energy $E_{\text{int}}$ equals to the kinetic energy in the long-time expansion and could be extracted by measuring atom expansion velocity. In Fig. \ref{fig:sFig1}(a), we measure $E_{\text{int}}$ at three typical modulation times, i.e., $t_{1}=6.6  \textrm{ ms}, \ 6.8  \textrm{ ms},\ 6.9  \textrm{ ms}$. The expansion velocity is determined experimentally by fitting the slope of the atomic cloud radius as a function of the time of flight (TOF). With the modulation time $t_{1}$ increasing from $6.6  \textrm{ ms}$ to $6.9  \textrm{ ms}$, the internal energies are determined as $E_{\text{int}}=0.84E_{0},\ 0.48E_{0},\ 0.28E_{0}$, respectively, where $E_{0}$ is the energy of the initial equilibrium system. 

The trapping potential energy at $t=t_{1}$ is determined by the in-situ cloud size
\begin{equation}
E_{\text{ho}}\left(t_{1}\right)=\frac{1}{2}m\omega_{1}^{2}\left\langle r^{2}\right\rangle \left(t_{1}\right).\label{eq:theory14}
\end{equation}
Since the in-situ cloud size $\left\langle r^{2}\right\rangle \left(t_{1}\right)$ is too small to be directly measured, it can be determined by probing atomic cloud after a TOF expansion. Specifically, we have the equation $\left\langle r^{2}\right\rangle \left(t_{1}\right)=\left\langle r^{2}\right\rangle \left(t_{1}+t_{\text{TOF}}\right)/b_{\text{TOF}}^{2}$, where the expansion factor $b_{\text{TOF}}$ satisfies the equation $\ddot{b}-\omega_{1}^{2}/b^{3}=0$ and the initial conditions of $b\left(t_{1}\right)$ and $\dot{b}\left(t_{1}\right)$ at time $t_{1}$ could be determined during the
excitation. Then we can calculate the trapping potential energy in the trap according to Eq. \eqref{eq:theory14}. Fig. \ref{fig:sFig1}(b) schematically shows the three in-situ atomic clouds from those in the TOF expansion with the correction of expansion factor. With the modulation time $t_{1}$ increasing from $6.6  \textrm{ ms}$ to $6.9  \textrm{ ms}$, the trapping potential energies are determined as $E_{\text{ho}}=0.28E_{0},\ 0.70E_{0},\ 1.00E_{0}$, respectively. This indicates that with the increase of trapping potential energy $E_{\text{ho}}$, the internal energy $E_{\text{int}}$ decreases.

\begin{figure} [htbp]
\centerline{\includegraphics[width=16cm]{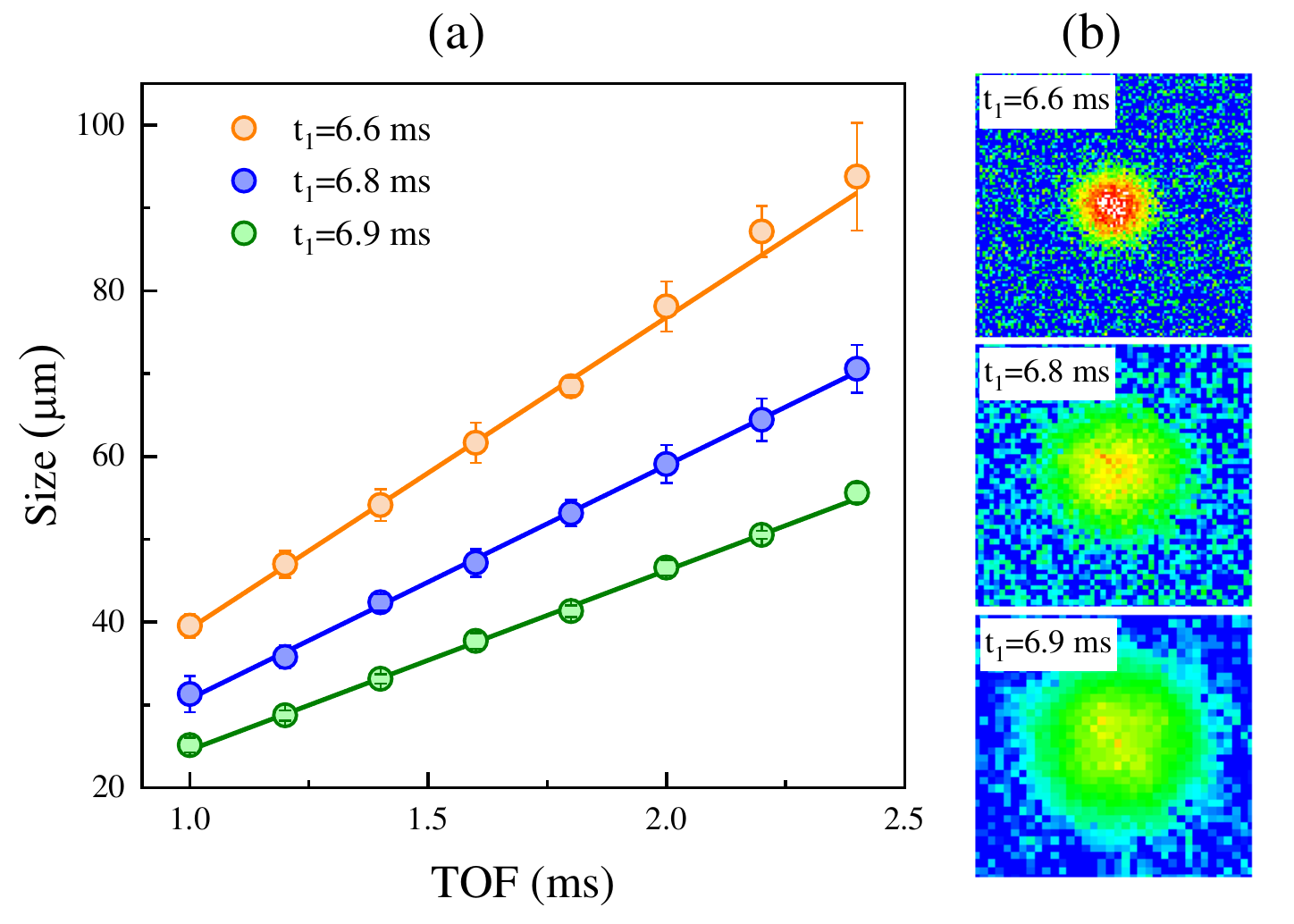}}
\caption{Measurements of internal energy $E_{\text{int}}$ and trapping potential energy $E_{\text{ho}}$. Three typical modulation times are at $t_{1}=6.6  \textrm{ ms}, \ 6.8  \textrm{ ms},\ 6.9  \textrm{ ms}$, respectively. (a) shows the atomic cloud sizes during the long-time expansion. From linearly fitting the cloud sizes (solid curves), we can get the atom expansion velocity and subsequently obtain the internal energy $E_{\text{int}}$. The error bar is the standard deviation of several measurements.  (b) schematically depicts the in-situ atomic clouds deduced from the absorption images after a TOF of 1 ms. Each image comes from one measurement. }
\label{fig:sFig1}
\end{figure}

\begin{figure} [htbp]
\centerline{\includegraphics[width=16cm]{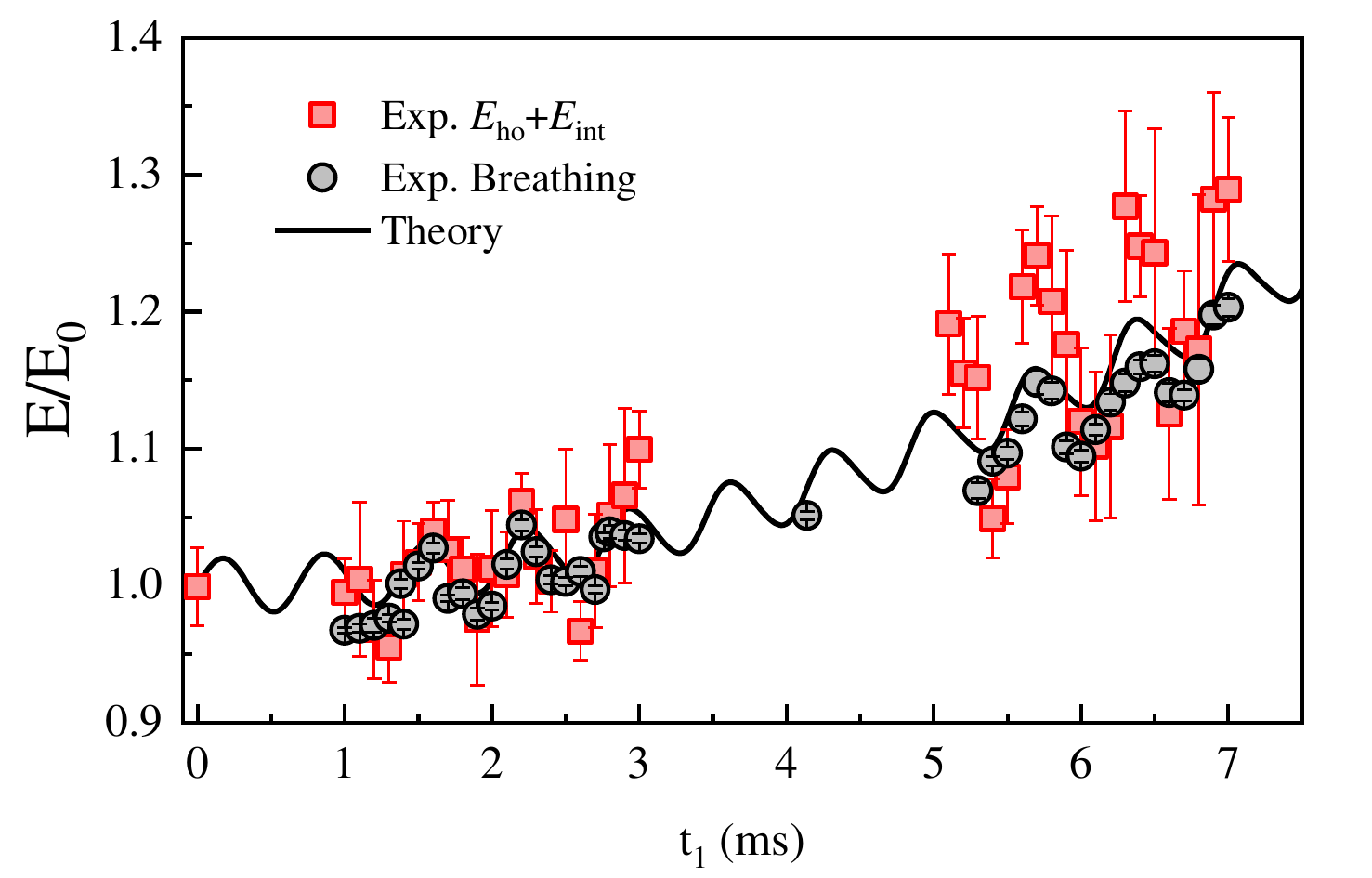}}
\caption{The system energy $E$ as a function of the modulation time $t_{1}$. The red squares denote the energy which is the sum of trapping potential energy $E_{\text{ho}}$ and internal energy $E_{\text{int}}$. The black circles represent the energy by measuring the breathing mode as in the main text.  The black solid curve denotes the calculation of Eq. \eqref{eq:theory8}. The modulation amplitude is $\beta=0.04$. The energy $E$ is normalized to the energy $E_{0}$ of the initial equilibrium system. The error bar is the standard deviation of several measurements.}
\label{fig:sFig2}
\end{figure}

The systematical measurements on $E_{\text{ho}}$ and $E_{\text{int}}$ are shown in Fig. 3 of the main text. There, $E_{\text{ho}}$ and $E_{\text{int}}$ oscillate nearly out of phase versus the manipulation time $t_{1}$, conversing between each other. To understand this behavior, let us consider an manipulation with a small amplitude $\beta\ll1$ and for a short time. During the manipulation, Eq. \eqref{eq:theory5} can be linearized by setting $b\left(t\right)\approx1+\epsilon\left(t\right)$ with $\epsilon\left(t\right)\ll1$. Then we obtain
\begin{equation}
\ddot{\epsilon}+4\omega_{0}^{2}\epsilon=-\beta\omega_{0}^{2}\sin\left(2\omega_{0}t\right)\label{eq:theory15}
\end{equation}
up to the first-order approximation of $\beta$ and $\epsilon\left(t\right)$, which admits the solution
\begin{equation}
\epsilon\left(t\right)=-\frac{\beta}{8}\left[\sin\left(2\omega_{0}t\right)-2\omega_{0}t\cos\left(2\omega_{0}t\right)\right].\label{eq:theory16}
\end{equation}
Thus we have 
\begin{equation}
b\left(t\right)\approx1-\frac{\beta}{8}\left[\sin\left(2\omega_{0}t\right)-2\omega_{0}t\cos\left(2\omega_{0}t\right)\right].\label{eq:theory17}
\end{equation}
The trapping potential energy and the internal energy are then given by
\begin{eqnarray}
\frac{E_{\text{ho}}(t)}{E_{0}} & = & \frac{1}{2}\frac{\omega^{2}\left(t\right)}{\omega_{0}^{2}}b^{2}\approx\frac{1}{2}+\frac{\beta}{4}\left[\frac{3}{2}\sin\left(2\omega_{0}t\right)+\omega_{0}t\cos\left(2\omega_{0}t\right)\right],\label{eq:theory18}\\
\frac{E_{\text{int}}(t)}{E_{0}} & = & \frac{E\left(t\right)-E_{\text{ho}}\left(t\right)}{E_{0}}\approx\frac{1}{2}+\frac{\beta}{4}\left[\frac{1}{2}\sin\left(2\omega_{0}t\right)-\omega_{0}t\cos\left(2\omega_{0}t\right)\right].\label{eq:theory19}
\end{eqnarray}
For $\omega_{0}t\gg3/2$ (or t$\gg0.3$ ms), which is fully satisfied within the timescale of the experiment, the cosine terms dominate in both $E_{\text{ho}}\left(t\right)$ and $E_{\text{int}}\left(t\right)$. Then Eqs. \eqref{eq:theory18} and \eqref{eq:theory19} are reduced as
\begin{equation} \label{eq:TwoComponentSpp}
\begin{split}
\frac{E_{\text{ho}}\left(t\right)}{E_{0}} & \approx  \frac{1}{2}+\frac{\beta}{4}\omega_{0}t\cos\left(2\omega_{0}t\right),\\
\frac{E_{\text{int}}\left(t\right)}{E_{0}} & \approx  \frac{1}{2}-\frac{\beta}{4}\omega_{0}t\cos\left(2\omega_{0}t\right).
\end{split}
\end{equation}
Equation \eqref{eq:TwoComponentSpp} clearly shows that $E_{\text{ho}}\left(t\right)$ and $E_{\text{int}}\left(t\right)$ oscillate nearly out of phase. 

The total energy could be determined as the sum of trapping potential energy and internal energy, i.e., $E=E_{\text{ho}}+E_{\text{int}}$. As mentioned above, $E_{\text{ho}}$ and $E_{\text{int}}$ are measured using the TOF expansion method. Together with the measurements from the breathing oscillation, the total energy $E$ is shown in Fig. \ref{fig:sFig2}. The measurements using the two methods are all consistent with the calculation of the dynamic virial theorem Eq. \eqref{eq:theory8}, while the experimental results using the TOF method have much larger fluctuations. This demonstrates that the long lifetime of breathing mode could strongly improve the energy-measurement accuracy.

\section{Effects of trap anharmonicity}

As the atomic cloud size increases, the anharmonicity of the trap becomes significant. This will affect the evolution of both the cloud size and energy. In this section, we discuss the effects of trap anharmonicity. The trapping potential experienced by atoms takes the form of
\begin{equation}
V_{\text{ext}}\left({\bf r},t\right)=U\left(t\right)\left[2-\left(e^{-2x^{2}/w^{2}}+e^{-2y^{2}/w^{2}}\right)e^{-z^{2}/w^{2}}\right],
\end{equation}
where $U\left(t\right)$ is the time-dependent potential at the trap center. To form a spherical optical dipole trap, the laser beams should satisfy the condition $w_{z}=\sqrt{2}w_{x,y}\equiv\sqrt{2}w$, where $w_{i}$ is the beam waist along the $i$-th direction \cite{wang2024PRLscaleSpp}. Expanding $V_{\text{ext}}\left({\bf r},t\right)$ near the trap center (${\bf r}=0$) up to the fourth-order terms in ${\bf r}$, we obtain
\begin{equation}
V_{\text{ext}}\left({\bf r},t\right)\approx\frac{1}{2}m\omega^{2}\left(t\right)\left\{ r^{2}-\frac{1}{2w^{2}}\left[r^{4}+\left(x^{2}-y^{2}\right)^{2}\right]\right\} ,
\end{equation}
where the trapping frequency is $\omega^{2}\left(t\right)=4U\left(t\right)/mw^{2}$. The collective oscillations can be excited by modulating the trapping potential, i.e., $U\left(t\right)=U_{0}\left[1+\beta\sin\left(\omega_{d}t\right)\right]$, where $\beta$ and $\omega_{d}$ are the modulation amplitude and frequency, respectively. Then the trapping frequency takes the form $\omega^{2}\left(t\right)=\omega^{2}_{0}\left[1+\beta\sin\left(\omega_{d}t\right)\right]$. The trapping frequency at equilibrium is $\omega_{0}=\sqrt{4U_{0}/mw^{2}}$. To make the quantities dimensionless, we introduce the length unit $a_{\text{ho}}=\sqrt{\hbar/m\omega_{0}}$ and energy $\hbar\omega_{0}$. We then have
\begin{equation}
V_{\text{ext}}\left({\bf r},t\right)\approx\frac{1}{2}\omega^{2}\left(t\right)\left\{ r^{2}-\frac{\eta}{2}\left[r^{4}+\left(x^{2}-y^{2}\right)^{2}\right]\right\} ,\label{eq:theory22}
\end{equation}
 where $\eta\equiv\left(a_{\text{ho}}/w\right)^{2}=\sqrt{\hbar^{2}/4mU_{0}w^{2}}$ characterizes the anharmonicity of the trap. To understand how the anharmonicity modifies the eigenmode frequencies, we set $\omega\left(t\right)=\omega_{0}$ and consider a small oscillation approximation, i.e., $b_{i}\approx1+\epsilon_{i}$ with $\epsilon_{i}\ll1$. Inserting Eq. \eqref{eq:theory22} into \eqref{eq:theory4}, we obtain
\begin{eqnarray}
\frac{d^{2}\epsilon_{x}}{dt^{2}}+2\left(\epsilon_{x}+\epsilon\right)-\frac{12\eta\chi_{0}}{5}\left(2\epsilon_{x}+\epsilon\right)-\frac{2\eta\chi_{0}}{5}\left(4\epsilon-\epsilon_{y}\right) & = & 0,\label{eq:theory23}\\
\frac{d^{2}\epsilon_{y}}{dt^{2}}+2\left(\epsilon_{y}+\epsilon\right)-\frac{12\eta\chi_{0}}{5}\left(2\epsilon_{y}+\epsilon\right)-\frac{12\eta\chi_{0}}{5}\left(4\epsilon-\epsilon_{x}\right) & = & 0,\label{eq:theory24}\\
\frac{d^{2}\epsilon_{z}}{dt^{2}}+2\left(\epsilon_{z}+\epsilon\right)-\frac{6\eta\chi_{0}}{5}\left(2\epsilon_{z}+\epsilon\right)-\frac{2\eta\chi_{0}}{5}\left(5\epsilon+\epsilon_{z}\right) & = & 0,\label{eq:theory25}
\end{eqnarray}
where $\epsilon\equiv\frac{1}{3}\sum_{i}\epsilon_{i}$, ($i \rightarrow x, y, z$). We have defined
\begin{equation}
\chi_{0}\equiv\frac{\int n_{0}\left(r\right)r^{6}dr}{\int n_{0}\left(r\right)r^{4}dr}
\end{equation}
and $n_{0}\left(r\right)$ is the equilibrium density profile and spherically symmetric to the first order in $\eta$. Solving these equations, we obtain modified eigen frequencies
\begin{eqnarray}
\omega_{1} & = & 2-\frac{19}{10}\eta\chi_{0},\label{eq:theory27}\\
\omega_{2} & = & \sqrt{2}-\frac{5}{3\sqrt{2}}\eta\chi_{0},\\
\omega_{3} & = & \sqrt{2}-\frac{13}{5\sqrt{2}}\eta\chi_{0},
\end{eqnarray}
where $\omega_{1}$ is for the breathing mode and $\omega_{2,3}$ for the quadrupole mode. In the limit $\eta\rightarrow0$, the results in a harmonic trap is recovered, i.e., $\omega_{1}=2$ and $\omega_{2,3}=\sqrt{2}$.

\section{Dynamic virial theorem in an anharmonic trap}

When anharmonicity is present, the dynamic virial theorem should be modified. Following a similar approach in Ref. \cite{peng2023PRAvirialSpp}, we obtain the relation
\begin{equation}
E\left(t\right)=\frac{1}{4}\frac{d^{2}I(t)}{dt^{2}}+\left\langle V_{\text{ext}}\right\rangle +\frac{1}{2}\left\langle {\bf r}\cdot\nabla V_{\text{ext}}\right\rangle ,
\end{equation}
which recovers the equilibrium result without trap modulation \cite{Thomas2008PRAvirialSpp},
\begin{equation}
E=\left\langle V_{\text{ext}}\right\rangle +\frac{1}{2}\left\langle {\bf r}\cdot\nabla V_{\text{ext}}\right\rangle .\label{eq:theory33}
\end{equation}
The damping of breathing oscillation is experimentally observed when the trap anharmonicity is large. To measure the energy injected into the system, the atoms are allowed to oscillate after the modulation and finally equilibrate in the trap, and then we probe atoms after a TOF expansion. The internal energy, $E_{\text{int}}=\left\langle {\bf r}\cdot\nabla V_{\text{ext}}\right\rangle /2$, is then determined 
by measuring the kinetic energy during the long-time expansion. Substituting the form of $V_{\text{ext}}\left({\bf r}\right)$ into the equilibrium virial theorem (\ref{eq:theory33}), we obtain
\begin{equation}
E\approx\left(2+\frac{19}{30}\eta\chi\right)E_{\text{int}}.
\end{equation}
Here, $\chi$ depends on the final temperature when the system equilibrates.


\begin{thebibliography}{40}
\bibitem [{Asp()}]{Asphericity}%
  \BibitemOpen
  \bibfield  {title} {\bibinfo {title} {{The} trap asphericity is defined as
  $\delta\equiv\left(\omega_{\text{max}}-\omega_{\text{min}}\right)/\bar{\omega}_{0}\approx5\%$,
  where $\omega_{\text{max}}$, $\omega_{\text{min}}$, $\bar{\omega}_{0}$ are
  the maximum, minimum and geometric mean frequencies along three axes,
  respectively},\ }\href@noop {} {\ }\BibitemShut {NoStop}%
\end{thebibliography}

%

\end{widetext}

\end{document}